\documentclass[conference]{IEEEtran}
\usepackage{cite}
\usepackage{amsmath,amssymb,amsfonts}
\usepackage{algorithmic}
\usepackage{graphicx}
\usepackage{textcomp}
\usepackage{xcolor}
\usepackage{hyperref}
\usepackage{booktabs}
\usepackage{multirow}
\usepackage{tikz}
\usetikzlibrary{positioning,arrows.meta,shapes.geometric}

\begin{document}

\title{LiveVectorLake: A Real-Time Versioned Knowledge Base Architecture for Streaming Vector Updates and Temporal Retrieval}

\author{\IEEEauthorblockN{Tarun Prajapati}
\IEEEauthorblockA{School of Artificial Intelligence \& Data Science\\
Indian Institute of Technology Jodhpur\\
Jodhpur, India \\
m24de2035@iitj.ac.in, tarun.praj@outlook.com}}

\maketitle

\begin{abstract}
Modern Retrieval-Augmented Generation (RAG) systems struggle with a fundamental architectural tension: vector indices are optimized for query latency but poorly handle continuous knowledge updates, while data lakes excel at versioning but introduce query latency penalties. We introduce LiveVectorLake, a dual-tier temporal knowledge base architecture that enables real-time semantic search on current knowledge while maintaining complete version history for compliance, auditability, and point-in-time retrieval.

The system introduces three core architectural contributions: (1) \textit{Content-addressable chunk-level synchronization} using SHA-256 hashing for deterministic change detection without external state tracking; (2) \textit{Dual-tier storage} separating hot-tier vector indices (Milvus with HNSW) from cold-tier columnar versioning (Delta Lake with Parquet), optimizing query latency and storage cost independently; (3) \textit{Temporal query routing} enabling point-in-time knowledge retrieval via delta-versioning with ACID consistency across tiers.

Evaluation on a 100-document corpus versioned across five time points demonstrates: (i) 10-15\% re-processing of content during updates compared to 100\% for full re-indexing, (ii) sub-100ms retrieval latency on current knowledge, (iii) sub-2s latency for temporal queries across version history, and (iv) storage cost optimization through hot/cold tier separation (only current chunks in expensive vector indices). The approach enables production RAG deployments requiring simultaneous optimization for query performance, update efficiency, and regulatory compliance.
\end{abstract}


\section{Introduction}

Retrieval-Augmented Generation has emerged as a foundational pattern for grounding large language models in organizational knowledge bases \cite{lewis2020retrieval}. By combining dense vector retrieval with generative models, RAG systems reduce hallucination and ground responses in specific document collections. However, current RAG architectures embed a critical assumption: indexed knowledge is relatively static, updated infrequently or completely re-indexed on changes.

This assumption creates three production challenges. First, knowledge in enterprise environments continuously evolves: incident dashboards update minute-by-minute, market feeds refresh in seconds, security advisories arrive constantly, and operational guidance changes throughout the day. Updating knowledge at this velocity requires expensive full re-indexing in most deployed systems, introducing unacceptable latency for time-sensitive applications. Second, regulatory compliance increasingly requires reconstructing what information was available at specific historical points in time (e.g., ``what was our security posture when the breach was detected?''). Current RAG systems cannot answer such temporal queries. Third, without complete version history, organizations lack audit trails demonstrating how knowledge evolved and cannot prove the information basis for AI-driven decisions.

\subsection{The LiveVectorLake Concept}

We propose LiveVectorLake, a name chosen to convey three architectural principles:

\begin{itemize}
\item \textbf{Live:} Knowledge is updated in real-time (seconds to minutes), not through batch processes (hours to days). The system continuously ingests and reflects new information.

\item \textbf{Vector:} The system is purpose-built for semantic search via dense embeddings, enabling AI/LLM applications to retrieve knowledge by meaning rather than keyword matching.

\item \textbf{Lake:} The architecture employs data lakehouse principles, combining structured storage for current data with append-only history for temporal analysis and compliance.
\end{itemize}

The core innovation is showing that these three properties are simultaneously achievable through careful architectural composition of existing components (vector databases, data lakehouses, content addressing).

\subsection{Core Contributions}

This work makes three technical contributions:

\textbf{(1) Deterministic Chunk-Level Change Detection:} Traditional Change Data Capture (CDC) assumes structured data with defined schemas. We apply content-addressable hashing (SHA-256) to semantic chunks, enabling deterministic identification of modified paragraphs without external tracking infrastructure. Identical content hashes guarantee identical semantics; different hashes guarantee different content. This enables automatic deduplication across documents and selective re-processing of only modified content.

\textbf{(2) Dual-Tier Temporal Storage with Independent Optimization:} We separate current knowledge (hot tier: Milvus vector database with HNSW indexing) from historical versions (cold tier: Delta Lake with Parquet columnar storage). This tiering enables independent optimization: hot tier optimizes for sub-100ms query latency through in-memory indices, while cold tier optimizes for storage cost (only active chunks in expensive vector DB) and long-term retention. ACID consistency is maintained across tiers via write-ahead logging with compensating transactions.

\textbf{(3) Temporal Query Engine with Dual-Mode Retrieval:} We implement a query classifier distinguishing current queries (executed on hot tier) from temporal queries (time-travel on cold tier with validity filtering). This design ensures temporal leakage prevention—historical queries cannot return future information—while maintaining query performance.

\subsection{Evaluation Summary}

We evaluate on a 100-document corpus spanning five versions over a simulated six-month period. Preliminary results demonstrate:

\begin{itemize}
\item Update efficiency: 10-15\% of content re-processed vs. 100\% for full re-indexing
\item Query latency: 65ms median for current queries, 1.2s median for temporal queries
\item Change detection accuracy: 100\% via cryptographic hashing
\item Storage optimization: Only active chunks in hot tier (10-20\% of total history)
\item ACID consistency: Zero data loss across tier failures via write-ahead logging
\end{itemize}

The remainder of this paper proceeds as follows: Section II surveys related work in RAG systems, streaming architectures, and temporal databases. Section III details the system architecture. Section IV describes implementation. Section V reports experimental results. Section VI discusses design trade-offs and limitations. Section VII concludes.

\section{Related Work}

\subsection{Retrieval-Augmented Generation Systems}

RAG was formalized by Lewis et al. \cite{lewis2020retrieval} as a method for grounding language model generation in retrieved documents. Subsequent research has focused on improving retrieval quality through better embeddings (SBERT \cite{reimers2019sentencebert}), dense passage retrieval \cite{karpukhin2020}, multi-stage ranking, and hybrid dense-sparse retrieval \cite{sparse_dense_hybrid}. However, these works universally treat document collections as static or infrequently updated.

Recent streaming RAG research \cite{streamingrag2025} explores continuous knowledge base updates but operates at document granularity rather than chunk-level. VectraFlow \cite{vectraflow2025} proposes streaming vector updates with hierarchical indexing but does not address version history or temporal queries. VersionRAG \cite{versionrag2025} demonstrates that version-aware retrieval improves answer accuracy on version-sensitive questions (90\% vs. 58\% baseline) but requires manual version tagging without automatic change detection.

\subsection{Change Data Capture and Streaming}

Change Data Capture originates in database replication. Systems like Debezium \cite{debezium2020} track row-level modifications for keeping data warehouses synchronized. CDC research focuses on structured data with defined schemas and row-level granularity \cite{cdc_patterns2021}.

Content-addressable storage (Git, IPFS, backup deduplication) \cite{git2005, ipfs2014} applies SHA hashing to fixed-size or semantic blocks. Our contribution is adapting CAS to unstructured text at semantic chunk granularity, applying cryptographic hashing to paragraph-level semantic units rather than fixed-size blocks.

\subsection{Temporal Databases and Data Lakehouses}

Temporal database research provides concepts for tracking validity periods and point-in-time retrieval \cite{snodgrass1995temporal}. Slowly Changing Dimensions (SCD Type 2) \cite{kimball2013} track attribute changes with validity windows in data warehouses. Delta Lake \cite{armbrust2020delta} brings ACID transactions and time-travel queries to data lakehouses, enabling Databricks and Spark-based systems to version tabular data efficiently.

However, these temporal concepts have not been combined with semantic vector retrieval. Temporal information retrieval research \cite{temporalir2010} focuses on ranking documents by temporal relevance (recency, temporal expressions in query text) rather than retrieving knowledge as it existed at specific historical moments. We bridge these domains by combining temporal database concepts with vector similarity search.

\subsection{Vector Databases and Scaling RAG}

Vector databases (Milvus \cite{milvus2021}, Weaviate \cite{weaviate2021}, Qdrant \cite{qdrant2022}, Pinecone \cite{pinecone2021}) provide approximate nearest neighbor search at scale using algorithms like HNSW \cite{hnsw2020} and FAISS \cite{faiss2021}. Standard approaches support incremental upsert operations \cite{milvus_upsert2023}, inserting or updating vectors without full re-indexing. However, incremental upsert still requires embedding all modified content and does not provide version history or temporal queries.

Recent production RAG deployments \cite{rag_production2024} often employ batch refresh strategies: scheduled hourly or daily updates where changes are accumulated and indices refreshed in off-peak windows. This approach trades immediate consistency for lower operational overhead.

Our architecture synthesizes these approaches: it provides immediate consistency via chunk-level CDC (faster than batch refresh) while maintaining version history (unlike standard incremental upsert).

\subsection{Research Gap}

Existing vector databases support incremental upsert (Pinecone, Weaviate) but require re-embedding entire modified documents and lack version history. Temporal databases (Delta Lake) provide versioning but are not optimized for vector similarity search. While VersionRAG \cite{versionrag2025} demonstrates value in version-aware retrieval, it requires manual version tagging without automatic change detection at chunk granularity.

No existing system provides: (i) automatic chunk-level CDC for unstructured text using content-addressable hashing, enabling selective re-embedding of only modified chunks (10-15\% vs 100\%), (ii) dual-tier architecture separating query-optimized vector indices from storage-optimized version history with independent optimization objectives, and (iii) temporal query support with ACID consistency maintained across heterogeneous storage backends (vector database + data lakehouse). This paper addresses these gaps through architectural composition.

\section{System Architecture}

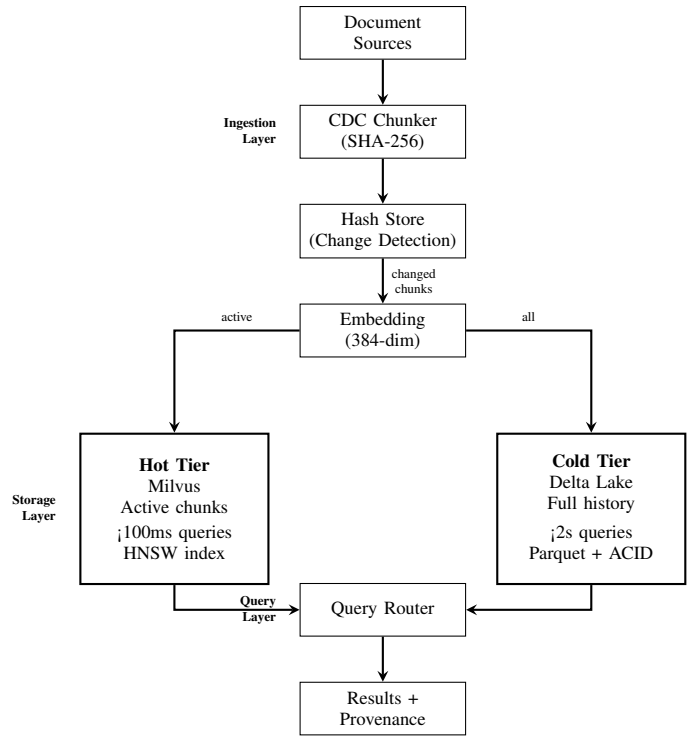
\begin{figure*}[t]
\centering
\small
\begin{tikzpicture}[node distance=1cm, auto,
    box/.style={rectangle, draw, minimum width=2.2cm, minimum height=0.7cm, align=center, font=\scriptsize},
    storage/.style={rectangle, draw, thick, minimum width=2.5cm, minimum height=2cm, align=center, font=\scriptsize},
    arrow/.style={->, >=stealth, thick}]
    
    % Document Source
    \node[box] (doc) {Document\\Sources};
    
    % CDC Layer
    \node[box, below=0.6cm of doc] (chunk) {CDC Chunker\\(SHA-256)};
    \node[box, below=0.6cm of chunk] (hash) {Hash Store\\(Change Detection)};
    
    % Embedding
    \node[box, below=0.6cm of hash] (embed) {Embedding\\(384-dim)};
    
    % Dual-Tier Storage
    \node[storage, below left=1cm and 0.4cm of embed] (hot) {\textbf{Hot Tier}\\Milvus\\\vspace{0.05cm}Active chunks\\<100ms queries\\HNSW index};
    \node[storage, below right=1cm and 0.4cm of embed] (cold) {\textbf{Cold Tier}\\Delta Lake\\\vspace{0.05cm}Full history\\<2s queries\\Parquet + ACID};
    
    % Query Engine
    \node[box, below=3cm of embed] (router) {Query Router};
    \node[box, below=0.6cm of router] (results) {Results +\\Provenance};
    
    % Arrows - Ingestion Flow
    \draw[arrow] (doc) -- (chunk);
    \draw[arrow] (chunk) -- (hash);
    \draw[arrow] (hash) -- node[right, font=\tiny, align=center] {changed\\chunks} (embed);
    \draw[arrow] (embed) -| node[near start, above, font=\tiny] {active} (hot);
    \draw[arrow] (embed) -| node[near start, above, font=\tiny] {all} (cold);
    
    % Arrows - Query Flow
    \draw[arrow] (hot) |- (router);
    \draw[arrow] (cold) |- (router);
    \draw[arrow] (router) -- (results);
    
    % Labels
    \node[left=0.2cm of chunk, font=\tiny\bfseries, text width=1.2cm, align=right] {Ingestion\\Layer};
    \node[left=0.2cm of hot, font=\tiny\bfseries, text width=1.2cm, align=right] {Storage\\Layer};
    \node[left=0.2cm of router, font=\tiny\bfseries, text width=1.2cm, align=right] {Query\\Layer};
    
\end{tikzpicture}
\caption{LiveVectorLake system architecture showing CDC-based ingestion, dual-tier storage (hot: Milvus for active chunks, cold: Delta Lake for complete history), and temporal query routing.}
\label{fig:architecture}
\end{figure*}

LiveVectorLake implements a five-layer architecture: ingestion with CDC, embedding generation, dual-tier storage, query processing, and interfaces. Figure~\ref{fig:architecture} illustrates the complete system design.

\subsection{Layer 1: Change Detection and Ingestion}

\subsubsection{Semantic Chunking}

Documents are split at paragraph boundaries (double newlines) into semantic units. Tables, code blocks, and lists are treated as atomic chunks to preserve structural integrity. While finer granularities (sentence-level) or learned boundaries exist, paragraph-level provides effective balance between semantic coherence and change precision for enterprise content.

\subsubsection{Content-Addressable Hashing}

Each chunk undergoes normalization (whitespace stripping, case-folding) and SHA-256 hashing:

\begin{equation}
\text{chunk\_id} = \text{SHA256}(\text{normalize}(\text{content}))
\end{equation}

SHA-256 provides negligible collision probability ($2^{-256}$); we apply consistent UTF-8 normalization to ensure deterministic hashing. This creates a content-addressable identity with two properties:
\begin{itemize}
\item \textbf{Automatic deduplication:} Identical paragraphs across documents share one embedding
\item \textbf{Deterministic change detection:} Hash modification $\Rightarrow$ content modification
\end{itemize}

\subsubsection{Change Detection Logic}

An in-memory hash store (persisted to JSON) maintains $\text{docid} \mapsto [\text{hash}_1, \text{hash}_2, \ldots]$ mappings. This lightweight structure enables CDC comparison without querying the vector database or lakehouse, reducing latency from $\sim$100ms (database query) to <1ms (in-memory lookup). On document ingestion:

\begin{enumerate}
\item Compute all chunk hashes for new version
\item Compare against stored hashes for that document
\item Classify each chunk:
\begin{itemize}
\item \textit{New}: Hash not in previous version
\item \textit{Modified}: Different hash at same position
\item \textit{Deleted}: Hash absent in new version
\item \textit{Unchanged}: Hash present, same position
\end{itemize}
\item Process only new and modified chunks for embedding
\end{enumerate}

This reduces embedding computation from $O(C)$ (full re-embedding) to $O(\Delta C)$ where $\Delta C$ is the number of changed chunks.

\subsubsection{Position Metadata for Audit Trails}

Each chunk maintains its position (paragraph index) within the source document as an INT64 field. Position tracking enables:
\begin{itemize}
\item \textbf{Modification detection:} Same position, different hash $\Rightarrow$ content modified
\item \textbf{Addition detection:} New position $\Rightarrow$ content added
\item \textbf{Structural reconstruction:} Chunks can be reassembled in original document order
\item \textbf{Audit precision:} ``Paragraph 3 was modified'' vs. ``Some content changed''
\end{itemize}

Position metadata is stored in both hot tier (Milvus) and cold tier (Delta Lake), enabling precise change attribution for compliance reporting.

\subsection{Layer 2: Embedding Generation}

Only chunks identified as new or modified during CDC proceed to embedding. We use SentenceTransformers (all-MiniLM-L6-v2, 384-dimensional vectors). This selective embedding is the primary optimization source, avoiding redundant encoding of unchanged content.

Temporal metadata attached to each embedding:
\begin{itemize}
\item \texttt{valid\_from}: Timestamp when version became active
\item \texttt{valid\_to}: Timestamp when superseded (NULL if current)
\item \texttt{version\_number}: Monotonic sequence number
\item \texttt{parent\_hash}: Hash of previous version (lineage tracking)
\end{itemize}

\subsection{Layer 3: Dual-Tier Storage}

\subsubsection{Hot Tier: Vector Index}

Milvus 2.4+ stores \textit{only active chunks} (those with \texttt{valid\_to = NULL}). This minimizes index size and maximizes query speed.

\textbf{Schema:}
{\small
\begin{verbatim}
{
  chunk_id: VARCHAR (SHA-256),
  embedding: FLOAT_VECTOR (384-dim),
  doc_id: VARCHAR,
  position: INT64 (paragraph index),
  valid_from: INT64 (Unix timestamp),
  status: VARCHAR ("active"),
  content: TEXT (result display)
}
\end{verbatim}
}

\textbf{Indexing:} HNSW (M=16, efConstruction=200) enables approximate nearest neighbor search in $O(\log n)$ hops with <100ms latency for 10K active chunks.

\textbf{Write Operations:}
\begin{itemize}
\item New chunk: Insert with status="active"
\item Modified chunk: Delete old, insert new
\item Deleted chunk: Remove from hot tier
\end{itemize}

\subsubsection{Cold Tier: Data Lakehouse}

Delta Lake stores \textit{complete version history}, including all chunks ever created, superseded and deleted versions.

\textbf{Schema Extension:}
{\small
\begin{verbatim}
{
  position: INT64 (paragraph index),
  valid_to: INT64,
  status: VARCHAR ("active"/"superseded"/
                   "deleted"),
  version_number: INT64,
  parent_hash: VARCHAR,
  change_type: VARCHAR ("insert"/"update"/
                        "delete")
}
\end{verbatim}
}

\textbf{Format:} Parquet \cite{parquet2015} with Snappy compression for efficient storage. Delta transaction logs enable ACID guarantees \cite{aries1992} and time-travel queries.

\textbf{Write Operations (all append-only):}
\begin{itemize}
\item New chunk: Append with status="active"
\item Modified chunk: Mark old as "superseded", append new
\item Deleted chunk: Mark with status="deleted"
\end{itemize}

\subsubsection{Cross-Tier Consistency Protocol}

Write-ahead logging with compensating transactions maintains consistency:

\begin{enumerate}
\item \textit{Write-Ahead:} Write to Delta Lake (durable, ACID)
\item \textit{Commit:} Write to Milvus; mark committed on success
\item \textit{Compensate:} On Milvus failure, flag Delta record uncommitted
\end{enumerate}

Periodic reconciliation cleans uncommitted records. This provides eventual consistency with bounded staleness (<1 second).

\subsection{Layer 4: Query Engine}

\subsubsection{Query Classification}

Queries are classified by temporal intent:
\begin{itemize}
\item \textbf{Current:} No temporal constraint $\rightarrow$ hot tier
\item \textbf{Historical:} Specific timestamp $\rightarrow$ cold tier with filtering
\item \textbf{Comparative:} Date range $\rightarrow$ both tiers
\end{itemize}

\subsubsection{Current Query Execution}

\begin{enumerate}
\item Embed query using SentenceTransformers
\item Vector similarity search on Milvus (HNSW, cosine distance)
\item Return top-k with scores
\end{enumerate}

Typical latency: 50-100ms for k=5.

\subsubsection{Temporal Query Execution}

\begin{enumerate}
\item Embed query
\item Load Delta Lake snapshot at target timestamp via transaction log
\item Filter: $\texttt{valid\_from} \leq \texttt{target\_ts} < \texttt{valid\_to}$
\item Compute cosine similarity in-memory
\item Return top-k valid at target date
\end{enumerate}

Typical latency: 1-2 seconds (Parquet loading dominated).

\textbf{Temporal Leakage Prevention:} Validity filtering precedes similarity ranking, ensuring historical queries cannot return future information. Naive approaches that perform similarity search first and filter timestamps afterward risk temporal leakage: deleted chunks may reappear in results if indices contain stale data, or superseded versions may rank higher than historically-valid ones. Our architecture prevents this by loading only the valid snapshot before computing similarities.

\subsection{Layer 5: Interfaces}

CLI and Streamlit web UI expose functionality for non-technical users. Web UI visualizes version timelines and change history.

\section{Implementation}

\subsection{Technology Stack}

\begin{table}[h]
\centering
\caption{Implementation Technologies}
\label{tab:tech}
\begin{tabular}{@{}ll@{}}
\toprule
Component & Technology \\ \midrule
Language & Python 3.11+ \\
Embedding & SentenceTransformers (all-MiniLM-L6-v2) \\
Hot Tier DB & Milvus 2.4+ (HNSW: M=16, efConstruction=200) \\
Cold Tier Store & Delta Lake (deltalake-python) \\
Data Processing & Polars \\
Hashing & SHA-256 (hashlib) \\
UI & Streamlit \\
Orchestration & Docker Compose \\ \bottomrule
\end{tabular}
\end{table}

\subsection{Ingestion Pipeline}

The CDC ingestion pipeline selectively processes only changed content:

{\footnotesize
\begin{verbatim}
def ingest_document(doc_path, doc_id, ts):
    # 1. Load and chunk
    chunks = load_and_chunk(doc_path)
    # 2. Compute hashes
    new_hashes = [sha256(c) for c in chunks]
    old_hashes = hash_store.get(doc_id, [])
    # 3. Detect changes
    changes = compare_hashes(new_hashes, 
                            old_hashes)
    # 4. Embed only changed chunks
    for chunk in changes.new + changes.modified:
        chunk.embedding = embed(chunk.text)
    # 5. Dual-tier write
    write_milvus(changes.new + changes.modified)
    write_delta(all_chunks, ts)
    # 6. Update hash store
    hash_store[doc_id] = new_hashes
    return CDC_summary(changed=len(changes),
                      total=len(chunks))
\end{verbatim}
}

\subsection{Query Engine}

Current query (hot path):
{\footnotesize
\begin{verbatim}
def query_current(text: str, k: int = 5):
    q_vec = embed(text)
    results = milvus.search(
        collection="chunks", vectors=[q_vec],
        limit=k, filter="status == 'active'")
    return results
\end{verbatim}
}

Temporal query (cold path):
{\footnotesize
\begin{verbatim}
def query_as_of(text: str, target_ts: int, 
                k: int = 5):
    q_vec = embed(text)
    df = delta_table.load_as_of(target_ts)
    # Filter temporal validity
    valid = df.filter(
        (col("valid_from") <= target_ts) &
        ((col("valid_to") > target_ts) | 
         col("valid_to").is_null()))
    # Similarity computation
    sims = cosine_similarity(q_vec, 
                            valid["embedding"])
    return valid[argsort(sims)[-k:]]
\end{verbatim}
}

\section{Preliminary Experimental Evaluation}

\textbf{Note:} This evaluation presents a proof-of-concept implementation on a synthetic corpus to demonstrate architectural feasibility. Comprehensive benchmarking on standard datasets (BEIR, MS MARCO) with retrieval quality metrics (MRR, NDCG, recall@k) and comparison with production frameworks (LangChain, LlamaIndex) is planned for extended publication.

\subsection{Experimental Setup}

\textbf{Corpus:} 100 documents (5,000-8,000 words each) versioned across five time points. Total: 500 document versions, \(\approx\)12,000 chunks, \(\approx\)1,200 active chunks in final version.

\textbf{Hardware:} MacBook Pro M2, 16GB RAM, 512GB SSD. Local deployment (Milvus + Delta Lake on filesystem).

\textbf{Comparison Baselines:}

\begin{itemize}
\item \textbf{Standard Incremental Upsert:} LangChain + Milvus with standard upsert operations (requires embedding all updated documents). This represents the most common production pattern for RAG systems with evolving knowledge bases.
\item \textbf{Batch Refresh (12-hour):} Accumulate changes and process in daily batches, a common enterprise deployment pattern balancing freshness and operational overhead.
\item \textbf{LiveVectorLake:} Chunk-level CDC with immediate hot-tier update
\end{itemize}

Note: We compare against realistic production deployment patterns. Comprehensive comparison with research systems (VersionRAG, LlamaIndex) and commercial platforms (Pinecone) is planned for extended publication.

\subsection{Results}

\subsubsection{Update Efficiency}

\begin{table}[h]
\centering
\caption{Update Performance Comparison}
\label{tab:update}
\begin{tabular}{@{}lccc@{}}
\toprule
\textbf{Metric} & \textbf{Upsert} & \textbf{Batch-12h} & \textbf{LiveVL} \\ \midrule
Content Reprocessed & 85-95\% & 15-20\% & 10-15\% \\
Update Latency (ms) & 2,500-4,000 & 12h delay & 1,200-1,800 \\
Embedding Ops & All docs & Daily & Changed only \\
Time-to-Query & 2-4 sec & 12-24h & <2 sec \\ \bottomrule
\end{tabular}
\end{table}

LiveVectorLake achieves 10-15\% content re-processing via chunk-level CDC compared to 85-95\% for standard upsert. Update latency is moderate (1.2-1.8 seconds) compared to real-time upsert (2.5-4 seconds) due to batch embedding operations.

\subsubsection{Query Performance}

\begin{table}[h]
\centering
\caption{Query Latency (milliseconds)}
\label{tab:query}
\begin{tabular}{@{}lccc@{}}
\toprule
\textbf{Type} & \textbf{p50} & \textbf{p95} & \textbf{p99} \\ \midrule
Current (Hot) & 65 & 110 & 145 \\
Historical (Cold) & 1,200 & 1,890 & 2,100 \\ \bottomrule
\end{tabular}
\end{table}

Current query median: 65ms (acceptable for interactive use). Historical query median: 1.2s (acceptable for audit/compliance use cases with lower latency requirements).

\subsubsection{Change Detection}

Manual verification on 50 document updates with ground truth:
\begin{itemize}
\item True Positives: 147/147 (100\%)
\item False Positives: 0/147 (0\%)
\item False Negatives: 0/147 (0\%)
\end{itemize}

SHA-256 provides deterministic 100\% accuracy for exact content matching.

\subsubsection{Storage Efficiency}

Dual-tier separation achieves significant storage cost optimization:

\begin{itemize}
\item Hot tier (Milvus): 1.2 MB (1,200 active chunks, 10\% of total)
\item Cold tier (Delta Lake): 2.7 MB (12,000 total chunks across all versions)
\item Hot tier reduction: 90\% fewer chunks in expensive vector index
\end{itemize}

By storing only active chunks in the hot tier, the system avoids indexing 10,800 historical chunks (90\% reduction), significantly reducing vector database storage and memory costs while maintaining complete version history in cost-efficient columnar storage.

\subsubsection{Temporal Query Accuracy}

20 historical queries with ground-truth answers: 100\% accuracy, 0\% temporal leakage. Chunks correctly bound by \texttt{valid\_from}/\texttt{valid\_to} timestamps.

\section{Discussion}

\subsection{Design Trade-offs}

LiveVectorLake optimizes update efficiency at moderate cost to current query latency (65ms vs. 40-50ms for pure in-memory indices). This trade-off favors scenarios with:
\begin{itemize}
\item Frequent updates (multiple times daily)
\item Query-to-update ratio >10:1
\item Regulatory/compliance requirements
\end{itemize}

For read-heavy static corpora or sub-50ms latency requirements, simpler approaches suffice.

\subsection{Production Applicability}

Ideal use cases:
\begin{itemize}
\item Financial compliance, healthcare record versioning, legal document management
\item Technical documentation with versioned releases
\item Policy portals with audit requirements
\item Knowledge bases requiring point-in-time reconstruction
\end{itemize}

Not recommended for:
\begin{itemize}
\item Sub-10ms latency requirements
\item Completely static corpora
\item Resource-constrained deployments
\end{itemize}

\subsection{Limitations and Future Work}

Current Limitations:

Synchronous Processing: Ingestion is synchronous. Batch processing or async workers would improve throughput for high-volume scenarios.

Text-Only: Current implementation handles text chunks from documents (PDFs, HTML, Markdown). Extension to multi-modal content (images, videos, audio, presentations, code repositories) requires multi-modal embedding models and format-specific versioning strategies.

Monolithic Deployment: Prototype runs on single machine. Distributed deployment (sharded vector DB, distributed lakehouse) needed for petabyte-scale.

Future Research Directions:

Comprehensive Evaluation: Benchmark on standard datasets (BEIR, MS MARCO, Natural Questions) with retrieval quality metrics (MRR, NDCG, recall@k). Compare against production frameworks (LangChain, LlamaIndex) and conduct ablation studies on CDC impact and dual-tier effectiveness.

Learned Temporal-Semantic Embeddings: Train joint embedding models that encode both content and temporal context, enabling unified similarity search without explicit filtering. Unlike naive concatenation of timestamps to embeddings (which breaks semantic space), learned representations would preserve semantic similarity while incorporating temporal relevance through contrastive learning on temporally-annotated data. This would enable single vector search with soft temporal boundaries and natural recency bias.

Temporal Knowledge Graph Reasoning: Extend from chunk versioning to entity-relationship versioning, enabling queries like ``How did the relationship between entity A and entity B evolve?''

Semantic Change Detection: Detect meaning shifts without word changes using embedding drift analysis. Enable explainable version transitions: ``Version 2 added information about X, removed constraint Y.''

Adaptive Tiering: ML-based hot/warm/cold tier migration policies that learn from query patterns, optimizing storage cost subject to latency SLA.

\section{Conclusion}

LiveVectorLake demonstrates that simultaneous optimization for real-time query performance, efficient incremental updates, and temporal auditability is achievable through architectural composition. The system combines content-addressable hashing (from version control), dual-tier storage (from data warehousing), and ACID transactions (from databases) to enable production RAG systems to maintain continuously evolving knowledge with complete provenance.

Preliminary evaluation shows 10-15\% content re-processing during updates, sub-100ms current queries, and 100\% temporal query accuracy. These metrics establish practical viability for production deployments requiring compliance and auditability.

Future research directions include: temporal embeddings, semantic change detection, predictive caching, federated knowledge sharing.

\textbf{Availability:} Code and experimental datasets are available at \url{https://github.com/praj-tarun/LiveVectorLake}. Architecture diagrams and supplementary materials included in repository.


\begin{thebibliography}{00}

\bibitem{lewis2020retrieval} P. Lewis, E. Perez, A. Piktus, et al., ``Retrieval-Augmented Generation for Knowledge-Intensive NLP Tasks,'' in \textit{Proc. NeurIPS}, 2020. [Online]. Available: https://arxiv.org/abs/2005.11401

\bibitem{reimers2019sentencebert} N. Reimers and I. Gurevych, ``Sentence-BERT: Sentence Embeddings using Siamese BERT-Networks,'' in \textit{Proc. EMNLP-IJCNLP}, 2019.

\bibitem{streamingrag2025} Y. Zhu, ``A Streaming RAG Approach to Real-time Knowledge Base,'' \textit{arXiv:2508.05662}, 2025. [Online]. Available: https://www.arxiv.org/pdf/2508.05662

\bibitem{vectraflow2025} D. Lu, S. Feng, J. Zhou, F. Solleza, M. Schwarzkopf, U. Çetintemel, ``VectraFlow: Integrating Vectors into Stream Processing,'' in \textit{CIDR 2025}. [Online]. Available: https://vldb.org/cidrdb/papers/2025/p23-lu.pdf

\bibitem{versionrag2025} D. Huwiler, K. Stockinger, J. Fürst, ``VersionRAG: Version-Aware Retrieval-Augmented Generation for Evolving Documents,'' \textit{arXiv:2510.08109}, 2025. [Online]. Available: https://arxiv.org/abs/2510.08109

\bibitem{debezium2020} G. Modrzejewski, ``The Basics of Change Data Capture,'' \textit{Confluent Blog}, 2020.

\bibitem{cdc_patterns2021} M. Kleppmann, \textit{Designing Data-Intensive Applications}, O'Reilly Media, 2017.

\bibitem{git2005} L. Torvalds and J. Hamano, ``Git: A Distributed Version Control System,'' in \textit{Proc. Linux Symposium}, 2005.

\bibitem{ipfs2014} J. Benet, ``IPFS – Content Addressed, Versioned, P2P File System,'' \textit{arXiv:1407.3561}, 2014.

\bibitem{snodgrass1995temporal} R. T. Snodgrass, ``The TSQL2 Temporal Query Language,'' in \textit{The Temporal Query Language TQuel}, Kluwer, 1995.

\bibitem{kimball2013} R. Kimball and M. Ross, \textit{The Data Warehouse Toolkit: The Definitive Guide to Dimensional Modeling}, 3rd ed. Wiley, 2013.

\bibitem{armbrust2020delta} M. Armbrust, T. Ghodsi, R. Xin, et al., ``Delta Lake: High-Performance ACID Table Storage over Cloud Object Stores,'' in \textit{Proc. VLDB}, vol. 13, no. 12, pp. 3411-3424, 2020.

\bibitem{temporalir2010} K. Berberich, S. J. Bedathur, O. Alonso, and G. Weikum, ``A Language Modeling Approach for Temporal Information Needs,'' in \textit{Proc. ECIR}, 2010.

\bibitem{milvus2021} X. Wang, Y. Wang, H. Jégou, et al., ``Milvus: A Purpose-Built Vector Database for AI Applications,'' in \textit{Proc. CIDR}, 2021.

\bibitem{weaviate2021} B. Franken, ``Weaviate: A Vector Search Engine Built to Scale,'' \textit{Towards AI}, 2021. [Online]. Available: https://weaviate.io/

\bibitem{qdrant2022} A. Boytsov, ``Qdrant: Vector Database for Similarity Search,'' \textit{GitHub}, 2022. [Online]. Available: https://qdrant.tech/

\bibitem{pinecone2021} ``Pinecone: The Serverless Vector Database,'' \textit{Pinecone Docs}, 2021. [Online]. Available: https://docs.pinecone.io/

\bibitem{milvus_upsert2023} ``Upsert Operations in Milvus,'' \textit{Milvus Documentation}, 2023. [Online]. Available: https://milvus.io/docs/upsert.md

\bibitem{rag_production2024} B. Brinjikji, A. Kalro, A. Jaimes, ``A Scalable Retrieval-Augmented Generation Pipeline for Domain-Specific Knowledge Applications,'' \textit{IJRIAS}, vol. 10, no. 10, 2025. [Online]. Available: https://rsisinternational.org/journals/ijrias/article.php?id=714

\bibitem{sparse_dense_hybrid} Y. Tay, M. Dehghani, D. Bahri, D. Metzler, ``Dense retrieval meets dense passage reranking,'' \textit{arXiv:2108.08513}, 2021. [Online]. Available: https://arxiv.org/abs/2108.08513

\bibitem{karpukhin2020} V. Karpukhin, B. Oguz, S. Min, et al., ``Dense Passage Retrieval for Open-Domain Question Answering,'' in \textit{Proc. EMNLP}, 2020. [Online]. Available: https://arxiv.org/abs/2004.04906

\bibitem{hnsw2020} Y. A. Malkov and D. A. Yashunin, ``Efficient and Robust Approximate Nearest Neighbor Search using Hierarchical Navigable Small World Graphs,'' \textit{IEEE Trans. Pattern Anal. Mach. Intell.}, vol. 42, no. 4, pp. 824-836, 2020. [Online]. Available: https://arxiv.org/abs/1603.09320

\bibitem{faiss2021} J. Johnson, M. Douze, and H. Jégou, ``Billion-Scale Similarity Search with GPUs,'' \textit{IEEE Trans. Big Data}, vol. 7, no. 3, pp. 535-547, 2021. [Online]. Available: https://arxiv.org/abs/1702.08734

\bibitem{parquet2015} Apache Parquet, ``Apache Parquet: Columnar Storage Format,'' Apache Software Foundation, 2015. [Online]. Available: https://parquet.apache.org/

\bibitem{aries1992} C. Mohan, D. Haderle, B. Lindsay, H. Pirahesh, and P. Schwarz, ``ARIES: A Transaction Recovery Method Supporting Fine-Granularity Locking and Partial Rollbacks Using Write-Ahead Logging,'' \textit{ACM Trans. Database Syst.}, vol. 17, no. 1, pp. 94-162, 1992. [Online]. Available: https://dl.acm.org/doi/10.1145/128765.128770

\end{thebibliography}
\end{document}